\documentclass[aps,prl,amssymb,twocolumn,superscriptaddress,showpacs]{revtex4}

\usepackage{graphicx}
\usepackage{dcolumn}
\usepackage{bm}
\usepackage{color}

\begin{document}

\newcommand{\beq}{\begin{equation}}
\newcommand{\eeq}{\end{equation}}
\newcommand{\barr}{\begin{eqnarray}}
\newcommand{\earr}{\end{eqnarray}}

\def\bra#1{\langle #1 |}
\def\ket#1{| #1 \rangle}
\def\sinc{\mathop{\text{sinc}}\nolimits}
\def\cV{\mathcal{V}}
\def\cH{\mathcal{H}}
\def\cT{\mathcal{T}}
\renewcommand{\Re}{\mathop{\text{Re}}\nolimits}
\newcommand{\tr}{\mathop{\text{Tr}}\nolimits}

\author{Paolo Facchi}
\affiliation{Dipartimento di Matematica, Universit\`a di Bari,
        I-70125  Bari, Italy}
\affiliation{INFN, Sezione di Bari, I-70126 Bari, Italy}
\author{Giuseppe Florio} \affiliation{Dipartimento di Fisica,
Universit\`a di Bari,
        I-70126  Bari, Italy}
\affiliation{INFN, Sezione di Bari, I-70126 Bari, Italy}
\author{Giorgio Parisi}
\affiliation{Dipartimento di Fisica, Universit\`{a} di Roma "La Sapienza", Piazzale Aldo Moro 2, 00185 Roma, Italy\\
Centre for Statistical Mechanics and Complexity (SMC), CNR-INFM, 00185 Roma,
Italy\\
INFN, Sezione di Roma,  00185 Roma,
Italy}
\author{Saverio Pascazio} \affiliation{Dipartimento di Fisica,
Universit\`a di Bari,
        I-70126  Bari, Italy}
\affiliation{INFN, Sezione di Bari, I-70126 Bari, Italy}

\title{Maximally multipartite entangled states}

\date{\today}

\begin{abstract}
We introduce the notion of  maximally multipartite entangled states
of  $n$ qubits as a generalization of the bipartite case. These pure
states have a bipartite entanglement that does not depend on the
bipartition and is maximal for all possible bipartitions. They are
solutions of a minimization problem. Examples for small $n$ are
investigated, both analytically and numerically.
\end{abstract}


\maketitle

The characterization of multipartite entanglement is no simple
matter. The \emph{bipartite} entanglement of a composed quantum
system \cite{woot} can be consistently defined and quantified in
terms of the entropy of entanglement or some physically equivalent
quantity. On the other hand, there is no unique way of
characterizing the
\emph{multipartite} entanglement. Different definitions often do
not agree with each other, because they adopt different strategies,
focus on different aspects and capture different features of this
quantum phenomenon \cite{multipart,mw,scott}.  There is a profound reason
behind this problem: the number of measures (i.e.\ real numbers)
needed to quantify multipartite entanglement grows exponentially
with the size of the system (e.g., the number of qubits). Therefore,
the definition of appropriate entanglement measures, able to
summarize the most salient global features of entanglement, can be
very difficult and their evaluation bear serious computational
difficulties. This difficulty is a characteristic trait of
complexity \cite{parisi} and entanglement is no exception in this
respect \cite{MMSZ,FFP}: we shall introduce here multipartite
entangled states that bear the symptoms of frustration.

The aim of this Letter is to explore ``maximally" multipartite
entangled states of $n$ qubits. These states, to be precisely
defined later, are maximally (bipartite) entangled for \emph{all}
possible bipartitions. The focus is therefore on the global,
partition-independent, features of entanglement. We will consider
only \emph{pure} states, the extension to mixed states being not
straightforward, due to well-known phenomena such as bound
entanglement \cite{boundent}.

We consider an ensemble $S=\{1,2,\dots, n\}$ of $n$ qubits in the Hilbert space
$\mathcal{H}= (\mathbb{C}^2)^{\otimes n}$, whose state is
\begin{equation}
|\psi\rangle = \sum_{k\in Z_2^n} z_k |k\rangle , \quad z_k \in
\mathbb{C}, \quad \sum_{k\in Z_2^n} {|z_k|}^2 =1,
\label{eq:genrandomx}
\end{equation}
where $k=(k_i)_{i\in S}$, with $k_i\in Z_2=\{0,1\}$, and
\begin{equation} \label{eq:ki} \ket{k}=\bigotimes_{i\in S}
\ket{k_i}, \qquad \ket{k_i}\in \mathbb{C}^2_i .
\end{equation}
In order to analyze the multipartite features of the entanglement
shared by the qubits, we proceed as follows. Consider a
bipartition $(A,\bar{A})$ of the system, made up of $n_A$ and
$n_{\bar{A}}$ qubits, respectively, where $A \subset S$ is a subset
of $n_A$ elements, $\bar{A}=S\backslash A$ its complement,
$n_A+n_{\bar{A}}=n$ and we will stipulate $n_A \leq n_{\bar{A}}$
with no loss of generality. The total Hilbert space is accordingly
factorized into
$\mathcal{H}=\mathcal{H}_A\otimes\mathcal{H}_{\bar{A}}$, with
$\mathcal{H}_A= \bigotimes_{i\in A} \mathbb{C}^2_i$, of dimensions
$N_A=2^{n_A}$ and $N_{\bar{A}}=2^{n_{\bar{A}}}$, respectively
($N_AN_{\bar{A}}=N$). As a measure of the
\emph{bipartite} entanglement between the two subsets, we consider
the purity of subsystem $A$
\begin{equation}
\pi_{A}=\tr_A\rho_{A}^2, \quad \rho_{A}=\tr_{\bar{A}}|\psi\rangle\langle\psi|, \label{eq:puritydef}
\end{equation}
$\tr_{X}$ being the partial trace over subsystem $X$. We notice that
$\pi_{A}=\pi_{\bar{A}}$ and
\beq \label{eq:purityconstraint}
1/N_A\le\pi_{A}\le 1.
\eeq
State (\ref{eq:genrandomx}) can be written accordingly to the
bipartition $(A,\bar{A})$ as
\begin{equation}
|\psi\rangle = \sum_{k\in Z_2^n} z_k
\ket{k^A}\otimes\ket{k^{\bar{A}}} ,
\label{eq:genrandomxbip}
\end{equation}
where
$\ket{k^A}=\bigotimes_{i\in A}\ket{k_i}\in\mathcal{H}_A$. By
plugging Eq.\ (\ref{eq:genrandomxbip}) into Eq.\
(\ref{eq:puritydef}) we obtain
\beq
\pi_{A}= \sum_{k_i \in Z_2^n} z_{k_1}
\bar{z}_{k_2}z_{k_3}\bar{z}_{k_4} \delta_{k_1^A, k_4^A}
\delta_{k_2^A,k_3^A} \delta_{k_1^{\bar{A}}, k_2^{\bar{A}}} \delta_{k_3^{\bar{A}}, k_4^{\bar{A}}}
\label{eq:puritygeneral}.
\eeq

Notice that for a \emph{given} bipartition it is very easy to
saturate the lower bound $1/N_A$ of (\ref{eq:puritygeneral}): one
looks for those maximally \emph{bipartite} entagled states that
yield a totally mixed state $\rho_A=\openone /N_A$. We will
generalize the above property by requiring maximal possible mixedness for
\emph{each} subsystem $A\subset S$, given the constraint that
the total system be in a pure state.
A state endowed with this property will be called a
\emph{maximally multipartite entangled state} (MMES).

In the most favorable case this means that every subsystem $A$
composed of $n_A \leq n/2$ qubits is in a totally mixed state
$\rho_A=\openone /N_A$ and thus $\pi_A=1/N_A$ (recall that
$\pi_A=\pi_{\bar{A}}$, thus when $n_A>n/2$ the above requirement
applies to $\bar{A}$). In fact, it is sufficient to look at maximal
subsystems of size $n_A=[n/2]$ ($[x]=$integer part of $x$), because
the density matrix of every smaller part $B\subset A$ would
automatically be of the sought form, $\rho_B=\tr_{\bar{B}\cap
A}\rho_A=\openone/N_B$. Therefore a \emph{perfect}  MMES  would be
maximally entangled for \emph{every} bipartition $(A,\bar{A})$ and
would be characterized by $\pi_A=1/N_A$ for all
\emph{balanced} bipartitions.
Observe that the requirement of maximal mixedness for $(A,\bar{A})$,
$\pi_A=1/N_A$, \emph{and} the analogous requirement for a different
balanced bipartition $(B,\bar{B})$ , with $B \neq A$, might not be
compatible with each other, so that, at variance with the bipartite
case, perfect MMES do not necessarily exist.

We define a MMES as a minimizer of what we shall call the
\emph{potential of multipartite entanglement}
\begin{equation}
\label{eq:pime}
\pi_{\mathrm{ME}}=
\left(\begin{array}{l}n
\\n_A\end{array}\!\!\right)^{-1}\sum_{|A|=n_A}\pi_A ,
\end{equation}
where
$n_A=[n/2]$. The above quantity is related to the (average) linear
entropy $S_L=\frac{N_A}{N_A-1} (1-\pi_{\mathrm{ME}})$ introduced in
\cite{scott}, that extends ideas put forward in \cite{mw}. See also \cite{parthasarathy}.
The quantity $\pi_{\mathrm{ME}}$ measures the average bipartite
entanglement over all possible balanced bipartition and thus
inherits property (\ref{eq:purityconstraint}), i.e.\
\beq \label{eq:pmeconstraint}
1/N_A\le\pi_{\mathrm{ME}}\le 1.
\eeq
The upper bound $\pi_{\mathrm{ME}}=1$ is attained by the fully
factorized states,  $z_k=\prod_{i\in S} \alpha_{k_i}^i$, with
$|\alpha_0^i|^2 + |\alpha_1^i|^2=1$. On the other hand, the lower
bound $\pi_{\mathrm{ME}}=1/N_A$, if attained,  would correspond to a
perfect MMES, maximally entangled for every bipartition. However, it
can happen that the requirements of maximal mixedness for different
bipartitions compete with each other. In such a case, the system is
frustrated and the minimum of the potential (\ref{eq:pime}) is
strictly larger than the lower bound in (\ref{eq:pmeconstraint}),
i.e.\ $\min
\pi_{\mathrm{ME}}>1/N_A$.
Since in such situation it may happen that different bipartitions
yield different values of $\pi_A$, our strategy will be to seek
those states among the minimizers that have the smallest variance.

This quest can be recast as an optimization problem: search for the
minimum of the cost function
\beq
\tilde{\pi}_{\mathrm{ME}}(\lambda)=
\pi_{\mathrm{ME}}+\lambda \sigma_{\mathrm{ME}} ,
\label{eq:costfunction}
\eeq
where $\lambda \ge 0$ is a Lagrange multiplier and
\begin{equation}
\label{eq:sigma}
\sigma_{\mathrm{ME}}^2=
\left(\begin{array}{l}n \\n_A\end{array}\!\!\right)^{-1} \sum_{|A|=n_A}
\left(\pi_A-\pi_{\mathrm{ME}}\right)^2
\end{equation}
is the variance of $\pi_{A}$ over all balanced bipartitions. Notice
that the introduction of $\lambda$ enables one to look for a
compromise between the minimal purity $\pi_{\mathrm{ME}}$ (maximal
average entanglement) and the minimal standard deviation
$\sigma_{\mathrm{ME}}$ of the distribution (maximally distributed
entanglement). In general, the solution of this optimization problem
completely defines a class of states with the maximal possible
entanglement (minimum purity), that is also well distributed, being
as insensitive as possible to the particular choice of the
bipartition.

If $\lambda\gg 1$ the minimization process will yield a very well
peaked distribution of $\pi_A$ around its average: entanglement will be
uniformly distributed, but this does not necessarily provide a MMES;
for example, a completely separable state has a (vanishing)
entanglement that is insensitive to the change of the bipartition
($\pi_{A}=1$ for all bipartitions). More interesting is the case
$\lambda=0$; indeed, a solution that minimizes the cost function
$\tilde{\pi}_{\mathrm{ME}}(0)$ will have an entanglement
distribution centered on the minimum of the potential $\pi_{\mathrm{ME}}$.
Therefore, if this minimum saturates the lower bound in
(\ref{eq:pmeconstraint}), the width $\sigma_{\mathrm{ME}}$ must
vanish. This would be our desideratum. However, it is known that, for $n\geq 8$, perfect MMES do not exist \cite{scott}. The general problem is therefore complicated. As a first step we set $\lambda=0$ in Eq.\ (\ref{eq:costfunction}), and focus on
the minimization of the potential $\pi_{\mathrm{ME}}\equiv
\tilde{\pi}_{\mathrm{ME}}(0)$. We shall tackle this problem both
analytically and numerically. By plugging (\ref{eq:puritygeneral})
into (\ref{eq:pime}) one gets after some combinatorics
\begin{eqnarray}
\pi_{\mathrm{ME}}
= \sum_{k_i \in Z_2^n} \Delta(k_1,k_2;k_3,k_4) z_{k_1}
z_{k_2}\bar{z}_{k_3}\bar{z}_{k_4} ,
\label{eq:pimeDelta}
\end{eqnarray}
with
\begin{eqnarray}
& & \Delta(k_1,k_2;l_1,l_2)
=\left(\!\!\begin{array}{l}n \\n_A\end{array}\!\!\right)^{-1}\!\!\!\!\!\!\sum_{|A|=n_A} \!\!\!\! \delta_{k_1^{A}, l_1^{A}} \delta_{k_2^{A},l_2^{A}}
\delta_{k_1^{\bar{A}}, l_2^{\bar{A}}} \delta_{k_2^{\bar{A}}, l_1^{\bar{A}}}
\nonumber\\
& & \qquad\quad = g(k_1\oplus l_1 \vee k_2 \oplus l_2, k_1 \oplus l_2 \vee k_2\oplus l_1),
\nonumber\\
& & g(a,b) = \left(\begin{array}{l}n \\n_A\end{array}\!\!\right)^{-1} \delta_{a\wedge b, 0} \left(\begin{array}{c}
     n-|a|-|b|    \\
      n_A-|a|
\end{array} \right) ,
\label{eq:Delta}
\end{eqnarray}
where $a\oplus b=(a_i + b_i\; \mathrm{mod}\; 2)_{i \in
S}$ is the XOR operation, $a\vee b=(a_i + b_i - a_i b_i)_{i\in S}$
the OR operation, $a\wedge b=(a_i b_i)_{i\in S}$ the AND
operation, and  $|a|=\sum_{i\in S} a_i$. Equations
(\ref{eq:pimeDelta})-(\ref{eq:Delta}) yield a closed expression
for the average purity, that is amenable to analytic and numerical
investigation.

In order to further simplify the problem,
in the following discussion we will replace $r_k=|z_k|$ in Eq.\
(\ref{eq:genrandomx}) with its mean value $1/\sqrt{N}$, and focus
on the states
\begin{equation}
|\psi\rangle = \frac{1}{\sqrt{N}}\sum_{k\in Z_2^n} e^{i
\varphi_k}|k\rangle. \label{eq:randomphase}
\end{equation}
Plugging Eq.\ (\ref{eq:randomphase}) into
Eq.\  (\ref{eq:puritygeneral}) we find
\begin{eqnarray}
& &\pi_{A}=\frac{N_A+N_{\bar{A}}-1}{N} \nonumber\\
& & +\frac{2}{N^2}\sum_{l\neq l',m \neq m'}\cos\left(
\varphi^p_{lm}-\varphi^p_{l'm}+\varphi^p_{l'm'}-\varphi^p_{lm'}\right), \quad
\label{eq:purityphase}
\end{eqnarray}
where $\varphi^p_{lm}=\varphi_{p^{-1}(l,m)}$, $p$ being a
permutation such that $A=\{p(1),p(2),\dots, p(n_A)\}$. This is an
interesting formula, that is worth discussing in detail:
i) first of all, if $\varphi_k=\sum_{i\in S}
\varphi_{k_i}^i$ in (\ref{eq:randomphase}), one obtains separable
states, that yield the maximum possible value $\pi_{A}=1$ for all
bipartitions: indeed the $N(N_A-1)(N_{\bar{A}}-1)/4$ cosines in the
summation in (\ref{eq:purityphase}) are all 1; ii) the first
addendum in the right-hand side corresponds to the average
entanglement of typical states \cite{FFP,aaa}. Thus, the combination
of phases in the summation can increase or reduce the value of the
purity with respect to the typical one (at fixed bipartition); iii)
in order to get a lower value of purity, one should look for
combinations of angles that tend to yield negative cosines. On the
other hand, it is also clear that for $n>2$ not all cosines can be
$-1$, as this would yield a purity smaller than $1/N_A$; iv) at
fixed bipartition $(A,\bar{A})$  it is always possible to find
combinations of cosines that saturate the lower bound $\pi_A=1/N_A$.
However, when plugged into (\ref{eq:pimeDelta}), the requirement
that this lower bound be saturated for \emph{every} $A$ might not be
satisfiable. This problem is in general $n$-dependent.

We now explicitly look at the simplest examples. For two qubits we
only have one bipartition and the potential of multipartite
entanglement reduces to
\beq
\pi_{\mathrm{ME}}^{(2)}=\frac{3}{4}+\frac{1}{4}\cos\left(
\varphi_{0}-\varphi_{1}-\varphi_{2}+\varphi_{3}\right), \label{eq:costfunction2}
\eeq
where the indices of the phases are expressed again in terms of
$k$, in decimal notation. The minimization of the potential consists
in solving the equation $\pi_{\mathrm{ME}}^{(2)}=1/2$ (remember that
$N_A=N_{\bar{A}}=2$). It is straightforward to obtain
$\varphi_{0}-\varphi_{1}-\varphi_{2}+\varphi_{3}=\pi$,
which yields the MMES
\beq
\ket {\psi_2}=\frac{1}{2}\left(e^{i\varphi_0}\ket
0+e^{i\varphi_1}\ket 1+e^{i\varphi_2}\ket
2-e^{i(-\varphi_0+\varphi_1+\varphi_2)}\ket 3\right).
\label{eq:optimal2}
\eeq
In this degenerate case, multipartite entanglement coincides with
bipartite entanglement, and this state is obviously equivalent, up to local
operations, to a Bell state.

For three qubits one must look for the solutions of
$\pi_{\mathrm{ME}}^{(3)}=1/2$, where
\begin{eqnarray}
\pi_{\mathrm{ME}}^{(3)}=\frac{5}{8}&+& \frac{1}{48}\sum_{p}\left[
\cos(\varphi_{p(0)}+\varphi_{p(7)}-\varphi_{p(1)}-\varphi_{p(6)})
\right.
\nonumber\\
&+& \cos(\varphi_{p(2)}+\varphi_{p(5)}-\varphi_{p(4)}-\varphi_{p(3)})
\nonumber\\
&+& 2 \cos(\varphi_{p(0)}+\varphi_{p(3)}-\varphi_{p(1)}-\varphi_{p(2)})
\nonumber\\
&+& \left. 2 \cos(\varphi_{p(7)}+\varphi_{p(4)}-\varphi_{p(6)}-\varphi_{p(5)})\right],
\end{eqnarray}
where the sum is over the $3$ cyclic permutations (of the bits).
A class of solutions is
\begin{eqnarray}
\ket {\psi_3}&=&\frac{1}{8}(e^{i\varphi_0}\ket
0+e^{i\varphi_1}\ket 1+e^{i\varphi_2}\ket
2-e^{i(-\varphi_0+\varphi_1+\varphi_2)}\ket 3 \nonumber \\
&+& e^{i\varphi_4}\ket4-e^{i(-\varphi_0+\varphi_1+\varphi_4)}\ket 5
+e^{-i\varphi_6}\ket 6\nonumber\\
&+&e^{i(-\varphi_0+\varphi_1+\varphi_6)}\ket 7)
\label{eq:optimal3}
\end{eqnarray}
and form a 5-dimensional submanifold.
This state includes the GHZ state \cite{ghz} as a particular case
and shares the same properties of the GHZ state for what concerns
concurrence and one-tangle.  Note that $\pi_{\mathrm{ME}}^{(3)}$
contains 12 cosines with different arguments, 6 of which are counted
twice. The solution proposed corresponds to 2 cosines = 1, 4 cosines
$=-\beta$, $4\times 2$ cosines $=-1$ and $2 \times 2$ cosines
$=\beta$, with
$\beta=\cos(\varphi_0-\varphi_2-\varphi_4+\varphi_6)$, that sum up
to  $2 - 4\beta -8 + 4\beta=-6$. In fact, there are 3 families of
solutions of the form (\ref{eq:randomphase}), corresponding to the 3
submanifolds $M_p=\{\varphi_i |
\varphi_{p(0)}+\varphi_{p(7)}-\varphi_{p(1)}-\varphi_{p(6)}=0,
\varphi_{p(2)}+\varphi_{p(5)}-\varphi_{p(4)}-\varphi_{p(3)}=0,
\varphi_{p(0)}+\varphi_{p(3)}-\varphi_{p(1)}-\varphi_{p(2)}=\pi\}$
with $p$ a cyclic permutation. All three classes yield
the same pattern of cosines in $\pi_{\mathrm{ME}}^{(3)}$ ($\beta$
being given by the corresponding permutation).

For a number of qubits larger than 3, we turned to a numerical
approach: we generated a typical state of the form
(\ref{eq:randomphase}) and numerically tackled the minimization
problem through different kinds of iterative algorithms (for a
review of numerical techniques and their implementation, see
\cite{optimizationrev}). We first used deterministic algorithms. In
general, we found that minimization is strongly dependent on the
initial conditions. Therefore, we sampled a large number of initial
states, in order to test the reliability of the solutions obtained.
Among others, the truncated Newton method gave us the best results
in terms of both reliability and speed. The use of stochastic
algorithms gave us comparable results. In both cases the existence
of a large number of degenerate (local and global) minima required
an accurate analysis.

For $n=4$ qubits, we numerically obtained $\min
\pi_{\mathrm{ME}}^{(4)}\simeq 0.333>1/4$ with
$\sigma_{\mathrm{ME}}\simeq10^{-4}$. If the requirement $|z_k|=
1/\sqrt{N}$ in Eq.\ (\ref{eq:genrandomx}) [and
(\ref{eq:randomphase})] is relaxed, one can make
$\sigma_{\mathrm{ME}}$  vanish. This is a first example of
frustration among the bipartitions, that prevents the existence of a
perfect MMES. It is curious that the requirement that purity be
minimal for all balanced bipartitions generate conflicts already for
$n=4$ qubits. This is consistent with results obtained by other authors \cite{sudbery,borras}.

\begin{figure}[t]
\centering
\includegraphics[width=0.48\textwidth]{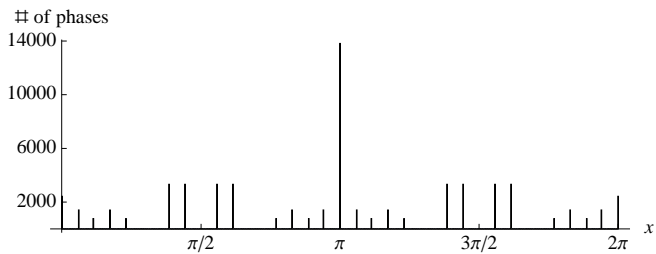}
\caption{
Distribution of the arguments $x$ of the cosines in
$\pi_{\mathrm{ME}}^{(6)}$, for a given 6-qubit MMES. Notice the
symmetries around $\pi$ and $\pi/2$. The total number of phases is
$62720$. Different phases are fully resolved by the binning.}
\label{6siti20run}
\end{figure}

For $n=5$ and 6, the landscape of the manifold on which the
minimization is performed becomes complicated. Nonetheless, we found perfect MMES, namely solutions for
$\pi_{\mathrm{ME}}^{(5)}=1/4$ and $\pi_{\mathrm{ME}}^{(6)}=1/8$,
respectively. Therefore, and curiously, frustration is present for
$n=4$ qubits, while it is absent for $n=5$ and 6. For example, a
5-qubits perfect MMES is defined by Eq.\ (\ref{eq:randomphase}) with
the following set of phases
\begin{eqnarray}
(\varphi_k)&=&(0,0,0,0,0,\pi,\pi,0, 0,\pi,\pi,0,0,0,0,0,\nonumber\\
&&0,0,\pi,\pi,0,\pi,0,\pi,\pi,0,\pi,0,\pi,\pi,0,0) \quad
\label{eq:ideal5}
\end{eqnarray}
and lives on a 7-dimensional manifold. The distribution of the
angles
$x=\varphi_{jl}^p-\varphi_{jl'}^p+\varphi_{j'l'}^p-\varphi_{jl'}^p$
for a given 6-qubit MMES is displayed in Figure
\ref{6siti20run}.
We observe interesting features, shared by all the MMES we
investigated and for all values of $n$: first of all, the
distribution is symmetric around $x=\pi$; second, there is a large
number of instances such that $\cos x=-1$, partially compensated by
the contribution of $\cos x=1$; third, the distribution of the
remaining angles is symmetric around $x=\pi/2$ and yields a
vanishing contribution to $\pi_{\mathrm{ME}}$.

For $n=7$ we numerically found configurations with $\min
\pi_{\mathrm{ME}}^{(7)}\simeq 0.134> 1/8$ and
$\sigma_{\mathrm{ME}}\simeq10^{-2}$, 
which improves  previous bounds \cite{scott,borras}.
By minimizing the cost function
(\ref{eq:costfunction}) with a non vanishing $\lambda$, one can reduce $\sigma_{\mathrm{ME}}$ to $\simeq10^{-3}$, at the expense of a higher $\pi_{\mathrm{ME}}\simeq 0.136$. It is not clear at present if the
impossibility to reach the absolute minimum for $n=7$ is to be
ascribed to the numerical procedure. 
For $n=8$ and $9$, where perfect MMES do not exist \cite{scott}, the convergence of the numerical simulations become very
slow. This is a typical signature of frustration.
The numerical results show that, in order
to obtain a vanishing width of the distribution for frustrated systems, it is necessary to increase the value of the
average of the purity. 
These conclusions are summarized in Table \ref{tab_mmes}.

A comment is now in order. Although in this article  perfect MMES are
exhibited as minimizers of the potential of entanglement, they are, by their very nature,  independent of the method used to find them. In fact, by virtue of their maximal mixedness, they saturate all measures of multipartite entanglement. This is the case, for
example, of the global entanglement measure of Meyer and Wallach
\cite{mw}, as well as its generalizations. Indeed, any entanglement
monotones, being functions of the partial density matrices of
subsystems of qubits, attain their maximal values on perfect MMES. The minimization procedure we propose is just a convenient way to construct them.
For non-perfect MMES, due to a possible finite value of $\sigma_{\mathrm{ME}}$, different measures of entanglement can resolve a part of the degeneracy of the manifold of minimizers.

It is interesting to briefly discuss one straightforward potential
application. Consider the 5-qubit perfect MMES $\ket{\psi_5}$
defined by Eqs.\ (\ref{eq:randomphase})  and (\ref{eq:ideal5}). One
can easily prove that
\barr
\bra{\psi_5} \sigma_1^z \sigma_2^z \sigma_3^y \sigma_4^y \sigma_5^z \ket{\psi_5}= 1 ,
\label{eq:appl5}
\earr
where $\sigma_i^y$ and $\sigma_i^z$ are Pauli matrices. Therefore,
the single- and two-qubit statistics are always ``flat", but the
measurements of the observables in (\ref{eq:appl5}) are always
strictly correlated.
\emph{Any} two parties, that can be far apart, can therefore share a
cryptographic key only if the other three parties agree on measuring
their respective observables in (\ref{eq:appl5}) and making their
results public. Notice that the key is shared by the two parties but
is unknown to the other three. We call this phenomenon
majority-agreed key distribution.

\begin{table}[t]
\caption{Perfect MMES for different $n$. }
\label{tab_mmes}
\begin{tabular}{|c|c|}
\hline
$n$ & perfect MMES \\
\hline
    2,3,5,6 & exist   \\
    4 & do not exist  \\
    7 & ?  \\
    $\geq 8$ & do not exist \\
    \hline
\end{tabular}
\end{table}

In conclusion, we introduced a class of multipartite entangled
states that maximize the amount and distribution of entanglement.
The features of these states strongly depend on the number of qubits
involved. In our numerical search, we noticed that already for a
relatively small number of qubits ($n \geq 7$), the landscape of the
parameter space where the optimization procedure is performed has a complex structure with a large number of local minima and a very slow convergence. 
The presence of frustration, due to the competition among different
partitions (observed already for $n=4$), appears to be a general feature of many-body
systems. It prevents the possibility to find perfect MMES but introduces interesting perspectives. In this sense the minimization task is a problem that requires a careful analysis  and the use of numerical and analytical strategies from different research fields. In this paper we just started to explore these connections.
Thus, the study of maximally multipartite entangled states paves the
way towards a deeper comprehension of the complex structure of quantum correlations arising in many-body systems.

\acknowledgments This work is partly supported
by the European Community
through the Integrated Project EuroSQIP.


\end{document}